\documentstyle[aps,prb,twocolumn,epsfig]{revtex}
\begin{document}
\twocolumn[\hsize\textwidth\columnwidth\hsize\csname
@twocolumnfalse\endcsname \draft
\title{Unoccupied electronic states of Au(113): \ theory and experiment.}
\author{Patricio H\"{a}berle*$^{1}$, Wladimir  Iba\~{n}ez$^{1}$, Rolando
Esparza$^{1}$ and Patricio  Vargas$^{1,2}$}

\address{{\it{ $^{1}$Departamento de F\'{\i}sica,
Universidad  T\'{e}cnica   Federico Santa  Mar\'{\i}a, Casilla
110-V,}\\ Valpara\'{\i}so, Chile}}

\address{{\it  $^{2}$Departamento de F\'{\i}sica,
Universidad  de  Santiago  de  Chile, P.O.Box 307,  Santiago-2,
Chile}}
\date{versi\'{o}n9.0, \today}
\maketitle

\begin{abstract}
We present results from Inverse photoemission spectroscopy in the
isochromat mode, with angular resolution, from the clean Au(113)
surface. \ To identify the origin of the different resonances we
have performed first principles calculations of the bulk band
structure in the LMTO\ formalism. \ Using the particular
characteristic of the spectrometer we have made a theoretical
prediction of the bulk features dispersion as a function of
parallel momentum, considering only energy and momentum
conservation. Thus we have been able to unambiguously identify,
from measured spectra various bulk derived resonances in addition
to two surface resonances and a surface state in the
$\lbrack\bar{1}10\rbrack$ and $\lbrack33\bar{2}\rbrack$ directions
respectively.

*email: phaberle@fis.utfsm.cl
\end{abstract}


]

\section{Introduction}
The electronic structure of low index faces of noble metals has
been the subject of many experimental studies using techniques
such as photoemission, two photon photoemission and inverse
photoemission. In particular several of those studies have dealt
with the electronic structure above the Fermi level
~\cite{wood82,reihl84Ag,bart85,Go85,smith85,chen89}
$(\varepsilon_{F})$ of different noble metal surfaces. The main
interest has been the description of image states and resonances
together with the identification of crystal derived surface
states~\cite{Fuggle}.

For Au(100), for example, there is data confirming the existence
of a surface state within the band gap at
$\bar{\Gamma}$~\cite{cicca94,straub86} and also a surface
resonance of a bulk derived features along the $\bar{X}
\bar{\Gamma}\bar{ M}$ directions. Similarly Au(111) also shows a
resonance which has been assigned to an image state in an energy
region above the band gap at $\bar{\Gamma}$
~\cite{straub86,wood86}.  On Au(110)~\cite{bart86} there are two
surface states at $\bar{X}$ and one at $\bar{Y}$ within a band
gap, for energies above $\varepsilon_{F}$. For Au surfaces then,
in every band gap at least one surface state has been detected;
image states have been observed, even if the states are within the
bulk allowed energy momentum region. All these surfaces have in
common that they show a room temperature reconstruction, but
little effect from it has been detected in the empty electronic
states. The results presented below are no exception to this
general rule.\\ Au is a still a subject of interest as a fairly
inert substrate to grow thin films of ferromagnetic
materials~\cite{himpsel91} that display oscillatory magnetization.
An important aspect, in these very thin films, is the mismatch
between the lattice parameters of the substrate and the film. Both
the morphology of the growth and therefore the physical properties
of the films are strongly dependent on this parameter. In the
search for the proper growth orientation and mass density of the
epitaxial layers, vicinal surfaces as fcc(113) could be
considered, but there is a lack of both experimental and
theoretical description of their electronic structure. In the case
of thin films~\cite{cram96,arena} both the width and intensity of
the unoccupied adsorbate induced resonances have been shown to
depend on the details of the substrate electronic structure.
\\

In the present study we describe the unoccupied electronic states
of Au(113) along the two principal axis of this surface.  We used
Inverse photoemission spectroscopy (IPS) together with first
principle calculations of the bulk band structure to provide a
complete interpretation of the origin and nature of the different
resonances present in our measurements. Using numerical
calculations to describe the bulk band structure and its
projection on a particular direction we have been able to label
the different surface resonances and states, independent of the
complexity of the measured surface electronic structure. This
numerical-experimental combination should prove valuable in the
description of the unoccupied states of thin metallic layers.

\bigskip
\section{Experimental}

Inverse photoemission spectroscopy (IPS) is a technique which
renders information regarding the unocuppied band structure of a
solid~\cite{pendry}.  The usual energy range considered goes from
$\varepsilon_{F}$ up to 10 or 15 eV above, including specially the
energy region below the vacuum level.  Our experiments were
performed in a vacuum chamber equipped with an isochromat inverse
photoemission spectrometer, based on a design by Dose
~\cite{den79srf}.  The photon detector is a Geiger M\"{u}ller
counter filled with Iodine as a discharge gas and He as a buffer
gas.  The window that accepts the photons into the detector is a
polished $SrF_{2}$ disc. The combination of the bandgap of the
window and the ionization potential of Iodine makes this detector
highly sensitive to photons in  a very narrow band around $(9.5
\pm 0.3)$ eV~\cite{Go85}. The electron beam is produced by an
electron gun based on a design by Zipf ~\cite{erdm82g} . It
consists of a BaO cathode indirectly heated by a tungsten
filament, an electron extraction element  and a focusing lens. The
measured energy resolution at 10 eV is 0.4 eV (FWHM) as measured
by detecting the current on a flat metallic sample, subject to a
ramp of increasing repulsive potential. The sample is mounted on a
goniometer with both an azimuthal rotation and a rotation through
an angle theta $(\theta)$ around  an axis on the plane of the
sample. This is an improved manipulator which allows a much more
precise and reproducible positioning than the one we used on a
preliminary measurement on this same system~\cite{hab95}. The
azimuthal angle is adjusted such that the electron momentum
parallel to the surface $(\hbar k_{//})$ is oriented along some
major crystallographic direction. By changing $\theta$ we can
change the angle between the surface normal and the incident
electronic momentum ($k$). Both the power supply which controls
the electron gun and the counter attached to the the detector are
connected to a personal computer via an interface using the GPIB
protocol. A typical spectrum shows the photon intensity as a
function of the energy of the incoming electrons in increments of
0.2 eV. The onset of the counts determines the location of
$(\varepsilon_{F})$.  A resonance in one of these spectrum can be
represented as a point in a energy-momentum $(\varepsilon$ vs.
$k_{//})$ plane using the relation $k_{//}
=sin\theta\sqrt{\frac{2m}{\hbar^{2}}(\varepsilon+\hbar\omega-\phi)}$,
with m being the electron rest mass, $\varepsilon$ the energy of
the resonance measured with respect to $(\varepsilon_{F})$,
 $\hbar\omega$ the energy of the detected photons, $\phi$ the work
function of the sample and $\theta$ has been defined above.

The sample was prepared from 5N Au boule which  was first
mechanically polished and then electroplolished with the surface
normal oriented within a 0.5$^{\circ}$ of the [113] direction as
verified by x-ray diffraction.  It was successively sputtered with
1 keV $Ar^{+}$ion beam and annealed to 450$^{\circ}$C.  The surface
displays a Low Energy Diffraction (LEED) pattern consistent with a
clean surface. It shows a reconstruction close to a (1$\times$5)
symmetry. ~\cite{sotto89,qtang}. A previous study of this surface
using x-ray diffraction~\cite{sandy} have shown this reconstruction
to be incommensurate with the substrate, but no structural model
for the surface has yet been proposed. From the LEED pattern itself
we can not determine if the phase is incommensurate unless a
detailed LEED I-V or other structural study is carried out.
\bigskip
\section{Numerical calculation}

In order to determine the origin of the  electronic resonances that
appear in our measurements. We performed a detailed calculation of
the pure bulk states of Au in the fcc structure at the equilibrium
lattice parameter.

The calculation of the electronic structure of  bulk Au was
performed using standard LMTO techniques~\cite{lmto}, with a
lattice parameter $a=4.08$\AA, in the fcc structure. A mesh of 18
points along each of the three primitive reciprocal translation
vectors and the tetrahedron method was used to perform integrations
in the Brillouin zone.
\begin{figure}[bt]
\centerline{\epsfig{file=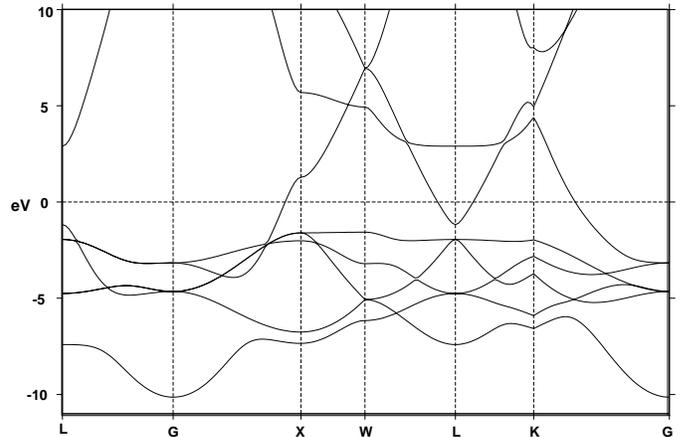,height=9cm,angle=-90}}
\caption{\label{Fig. 1.} Dispersion of the energy bands of Au
along main crystallographic directions as calculated in the LMTO
formalism }
\end{figure}

\begin{figure}[bt]
\centerline{\epsfig{file=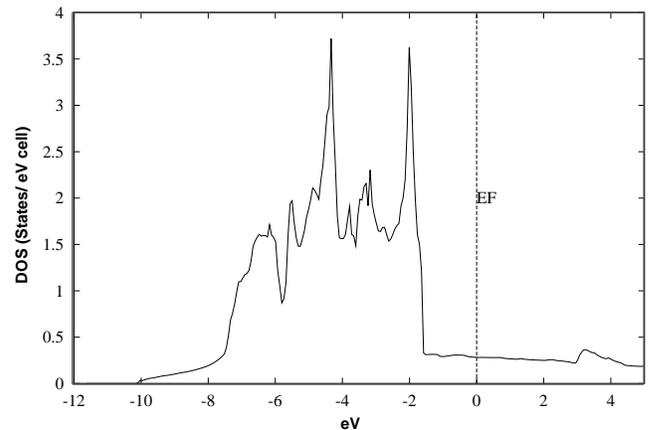,height=9cm,angle=-90}}
\caption{\label{Fig. 2.} The corresponding density of states for
the bands described in the previous figure.  The zero in the energy
axis is fixed at $(\varepsilon_{F})$.  The wide peak centered at
-4.5 eV corresponds to the d-band.  The states above
$(\varepsilon_{F})$ have a dominant s-p character. }
\end{figure}

 Figure 1 shows the result of
our calculation for the standard band structure in different
directions along the principal axis of the Brillouin zone. We can
also determine the total density of states (DOS) as a function of
energy (Figure 2) which could be compared to previous
calculations. In order to help us in the interpretation of our
experimental IPS measurements,we have performed a projection of
the electronic states along the two main perpendicular directions
of the (113) surface, namely $[\bar{1}10]$ and $[33\bar{2}]$
directions.  This operation is simply to represent in a single
graph all the energy states with a common $k_{//}$, regardless of
the momentum in the direction normal to the surface.   Figures 4
and 5 show the projected bands along the two perpendicular
directions (in units of \AA$^{-1}$) referred to the Fermi energy
(in $eV$). The density of points in the graph reminds us of the
underlying symmetry of the projected states which can be
visualized because for both figures we have used a mesh of 100
points for the complete range of $k_{\bot}$. It is easy to
recognize the existence of energy gaps between 1 eV and 3 eV above
the Fermi energy in both directions.   These gaps can be labeled
with a combination of X and L character, with $k_{X}=(1,0,0)$ and
$k_{L}=(\frac{1}{2},\frac{1}{2},\frac{1}{2})$ in units of
$2\pi/a$.

\begin{figure}[bt]
\centerline{\epsfig{file=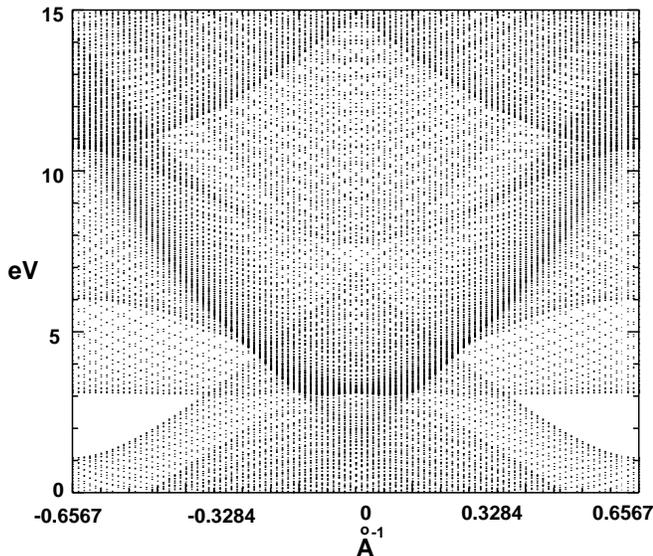,width=9cm}} \caption{\label{Fig.
3.} Au fcc bulk band structure projected in the (113) surface
along the $[33\bar{2}])$ direction in k space. The energy scale
are referred to $(\varepsilon_{F})$. The k axis is in \AA$^{-1}$
($a =4.08$ \AA). At the extreme of the SBZ we can see an energy
band gap between 1$eV$ and 3$eV$, which becomes narrower for small
$k_{//}$ until it disappears at $k_{//}\approx 0.33$\AA$^{-1}$.}
\end{figure}

To further visualize the location these energy gaps we calculated
all the electronic states on the unitary cube of side ($4\pi /a$)
in the reciprocal space. Figure 5 shows a constant energy surface
of the electronic states between 2.5 eV and 3 eV above
$\varepsilon_{F}$ in this unitary cube.  This small energy range
is chosen to provide enough points for a suitable representation
of the surface, as required by the smoothing and fitting routines
used to generate the graph. The cube has been rotated in such a
way that the [113] direction comes out normal to the plane of the
figure.

This constant energy surface shows two sets of gaps.  Each one of
them located symmetrically, along the two main perpendicular axis,
and also to the projection of the origin of the inverse space onto
the $(113)$ plane ($\bar{\Gamma}$).

\begin{figure}[bt]
\centerline{\epsfig{file=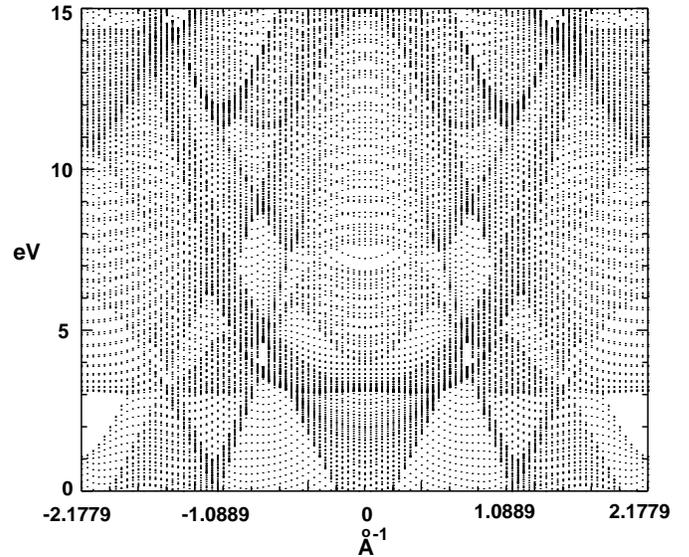,width=9cm}} \caption{\label{Fig.
4. } Au fcc bulk band structure projected in the $(113)$ surface
along the $[\bar{1}10]$ direction . The energy scale are referred
to $(\varepsilon_{F})$. The same band gap observed in the previous
figure can also be reached along this direction. }
\end{figure}
\begin{figure}
\centerline{\epsfig{file=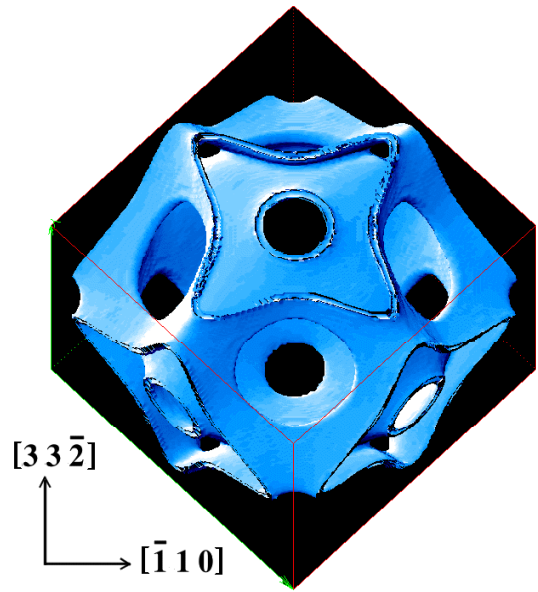,width=7cm}} \caption{\label{Fig.
5.}$k$-space surface of constant energy (2.5 $\rightarrow$ 3.0)eV
above $(\varepsilon_{F})$ for Au  electronic bulk states. The
$[$113$]$ direction in the reciprocal lattice is normal to the
plane of the figure. $\bar{\Gamma}$, the projection of origin in
k-space is at the symmetry center of the constant energy surface.
Two set of band gaps are clearly seen for the main perpendicular
directions. Only one of them retain this character (along
the[$33\bar{2}$] direction) after the electronic states are
projected onto the surface.}
\end{figure}

In an extended
representation of the inverse space only the gaps shown along the
$[33\bar{2}]$ direction remains. The rest of the space is filled by
the electronic states from other zones with their respective
centers slightly displaced. In particular the gaps along the
$[\bar{1}10]$ direction disappear and they can not be observed in
the surface projected band structure (see Figure 4).

The  two  gaps, located symmetrically with respect to
$\bar{\Gamma}$, along the $[33\bar{2}]$,could also be seen along
the perpendicular direction as shown Figures 3 and 4, simply
because of the peculiar shape of surface Brillouin zone for the
(113)surface (see Figure 6).

By inspection of Figure 5 we can clearly see that the gaps in this
energy region occurs by the exact superposition in k-space of the
projections of the necks joining the neighboring cubes along the
six [100] equivalent directions and the gaps in the zone boundary
along the cube diagonals ([111] direction). In the next section we
will see how  this information is relevant in the labeling of the
different surface resonances as seen by IPS. \bigskip

\begin{figure}
\centerline{\epsfig{file=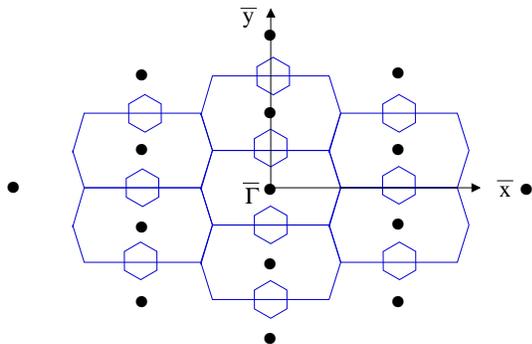,height=7cm,angle=90}}
\caption{\label{Fig. 6. }Representation of the SBZ for the (1x1)
fcc(113) surface.  The dots are the proyection of the reciprocal
space in the (113) plane. $\bar{x}$ and $\bar{y}$ are
$\lbrack\bar{1}10\rbrack$ and $\lbrack33\bar{2}\rbrack$ directions
respectively.  The hexagons represent the  location of the energy
gaps above the Fermi level in k-space. Starting from
$\bar{\Gamma}$ and going along $\bar{x}$ there is an energy gap
beyond the SBZ boundary. This same gap can also be reached along
the $\bar{\Gamma}-\bar{y}$ direction.  The projected electronic
structure of figures 3 and 4 show the actual extent of this gap in
each direction .}
\end{figure}

\bigskip
\section{Experimental results and discussion}

\subsection{\textbf{[$\bar{1}10$] direction}}
We will consider first a set of spectra along the the direction of
the close packed rows ($[\bar{1}10]$).  Since the reconstruction
of the surface shows no change in the surface periodicity along
this direction, as judged from the LEED diagrams, one should not
expect a large influence of the atomic rearrangement on the
surface electronic structure.  Figure 7 shows a series of IPS
spectra for different angles of the incoming electrons respect to
the surface normal.  The Fermi level is clearly distinguishable as
the onset for the photon intensity and it has been used as the
origin for the energy scale.  The intensity is measured as
photons/(electrons $\times$ energy) but they are presented in an
arbitrary scale. Some of the spectra have been re-scaled to
facilitate their display. \\ There are clearly two sets of
resonances, one them located above 10 $eV$ from $\varepsilon_{F}$
and the other one below 5 $eV$.  The high energy resonances are
fairly weak in intensity and show very little dispersion in
energy.  The low energy resonances instead are fairly well
defined, which makes  easier the identification of their evolution
as the angle respect to the surface normal is changed. Also these
states disperse over a wider energy range as can be seen directly
from Figure 7.

\begin{figure}[bt]
\centerline{\epsfig{file=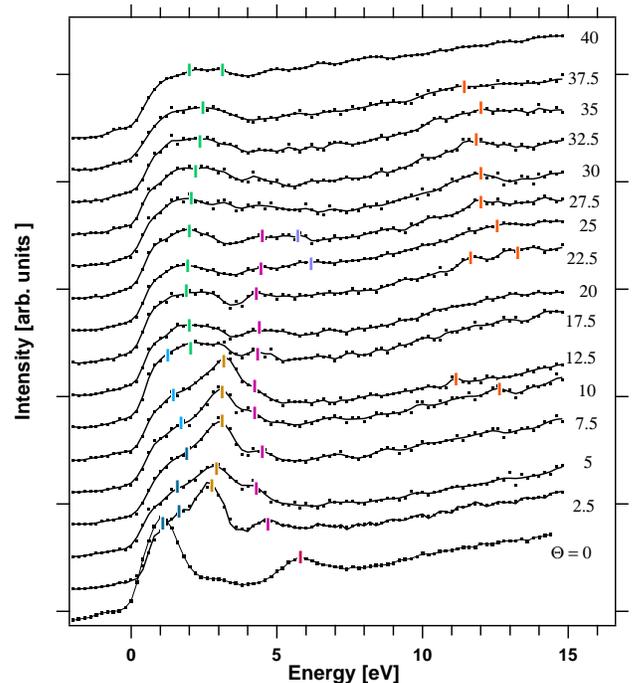,width=9cm}} \caption{\label{Fig.
7.} Superposition of a set of IPS spectra. They display the photon
intensity as a function of the energy of the incoming electrons.
The square data markers correspond to the actual reading of each
energy channel. The thin line is a smoothed spline to guide the
eyes through the data.  Several surface resonances have been
highlighted with vertical lines }
\end{figure}

In Figure 8 we show a plot of the  ${\varepsilon}$ vs. $k_{//}$
plane for this particular azimuth, with the parallel momentum
along the $[\bar{1}10]$ direction.  The filled squares are the
result of the numerical calculation.  They represent the energy
and momentum of a final state for a transition from an energy
state 9.5$\pm$ 0.3 $eV$ higher but for the same value of $k$.  The
calculated points are derived from the bulk energy bands
calculated as described in the previous section. This way of
finding the transitions is a more realistic than using the
parabolic approximation to map the energy dispersion of a
particular surface state or resonance.  The choice of the energy
difference is done to match the IPS's detector response.  In this
way, the square points correspond to a theoretical prediction of
bulk derived features of an IP experiment, based on a first
principle calculation of the solid energy bands. It should be also
noted that the energy difference between final and initial state
is not exactly 9.5 $eV$, we have added a 0.3 $eV$ Gaussian noise
to mimic the experimental response. This calculation does not
include  the spectral weight associated with the matrix element
effects on the optical transitions, which can  drive some of the
predicted events below the detectability limit.

\begin{figure}[bt]
\centerline{\epsfig{file=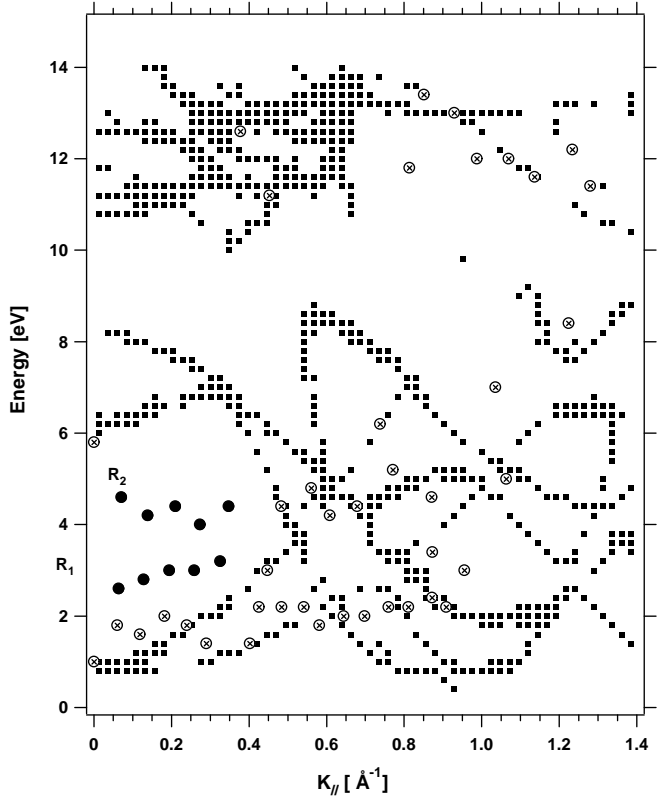,width=10cm}} \caption{\label{Fig.
8.} Dispersion of the surface resonances as a function of  $
k_{//}$. The experimental data is shown by the circles($\oplus$ and
$\bullet$) The theoretical prediction for bulk derived transitions
are labeled by the filled squares. $R_{1}$ and $R_{2}$ are surface
resonances. See text for further details.}
\end{figure}

The circular data markers in Figure 8, come from the measured data
(Figure 7) and they correspond to the different resonances in each
spectrum. We have chosen to separate them in two groups, one of
them, all data points that superpose  with calculated values of
bulk derived features$(\oplus)$ and the other set are those clearly
located in a gap$(\bullet)$ for the allowed bulk transitions.  It
should be clear that there is no absolute gap along this azimuth,
the voids in the $\varepsilon-k$ plane, shown in Figure 8 are
related to the restrictions on energy and momentum imposed on the
bulk transitions. We can then associate several resonances to bulk
derived features, but at the same time we can recognize the
existence of surface resonances, $(R_{1}$ and $R_{2})$ since they
show no superposition with the calculated transitions. $R_{1}$
behaves as a typical image state, since the energy is a minimum at
low $ k_{//}$ values and then it increases for larger $k$.
Rigorously it can not be labeled as an image state since it does
not happen in a gap of the bulk energy bands. The second resonance
$R_{2}$ at about 4.4 $eV$ above $\varepsilon_{F}$, just below the
vacuum level, could then be interpreted as the higher energy states
of the Rydberg series of image states.   It is indeed surprising
that these transitions are intense enough to be detected since
there is no absolute gap in the bulk states for this energy region,
hence the associated wave function should not be well localized at
the surface. The argument in favor for this interpretation in the
case of $R_{2}$, is the almost flat dispersion with $ k_{//}$ the
resonance, which keeps keeps it confined by the image potential
below the vacuum level.
\smallskip

\subsection {\textbf{$[33\bar{2}]$ azimuth}}
As in the perpendicular direction (Fig. 7) in Figure 9 we show a
complete set IPS spectra, where we can clearly identify several
resonaces with large dispersions, spanning in some cases an energy
range as large as 2 $eV$.
\begin{figure}[bt]
\centerline{\epsfig{file=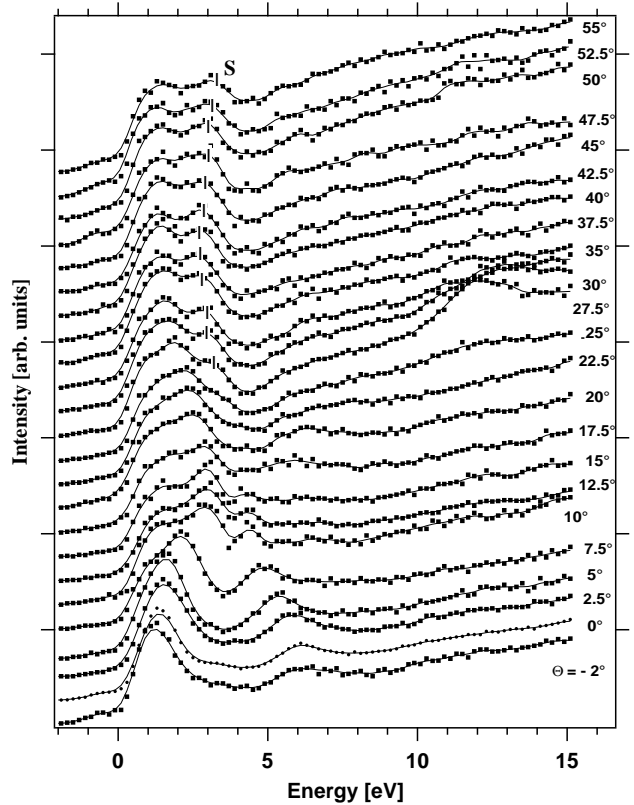,width=9cm}} \caption{\label{Fig.
9.} IPS spectra along the $\lbrack33\bar{2}\rbrack$ direction.
Measured data points are the solid squares. The resonance labeled
by $S$ is a surface state.}
\end{figure}
At low angle we have two prominent resonances in the spectra. One
of them which at normal incidence starts dispersing towards the
Fermi level from about 6 $eV$, down to 4$eV$ at about 15$^{\circ}$
off normal. Another prominent resonance starts from 1.3 $eV$ above
the Fermi level at normal incidence increasing its energy up to
2.9 $eV$ at about $17.5^{\circ}$.  At $30^{\circ}$ a new resonance
emerges ($SS$) at about $3$ $eV$ and as the angle increases it
moves towards the Fermi level arriving at a minimum energy at
about $40^{\circ}$.For larger angles it moves back to higher
energies and can it be clearly detected as far as $55^{\circ}$ off
normal.

We have represented all these resonances in a $\varepsilon$ vs.
$k$ plane in Figure 10.   The dark circular markers are the data
points taken from the IPS spectra and the black squares are again
the theoretical prediction for the bulk allowed transition. In
addition we have encircled in a solid line the region of the
absolute energy gap for this azimuth as determined from our
calculation shown in Figure 4.
 Most of the experimental data points are on top of the bulk
allowed transition and follow closely the dispersion of the
calculated features.  Exception to this statement is the state
labeled $S$, which is clearly contained in the absolute energy
gap, so we can label it legitimately as a surface state.  The
differences between the nature of $S$ and the resonances along the
$[\bar{1}10]$ direction, $R_{1}$ and $R_{2}$, are subtle but
clear. $R_{1}$ is contained in a "spectrometer energy gap", while
$S$ is within an absolute band gap. In addition $R_{1}$ shows a
typical upwards dispersion as $k_{//}$ increases. This behavior
could be associated with a free particle wave function dispersion
not necessarily derived from the crystalline energy bands.
\begin{figure}[bt]
\centerline{\epsfig{file=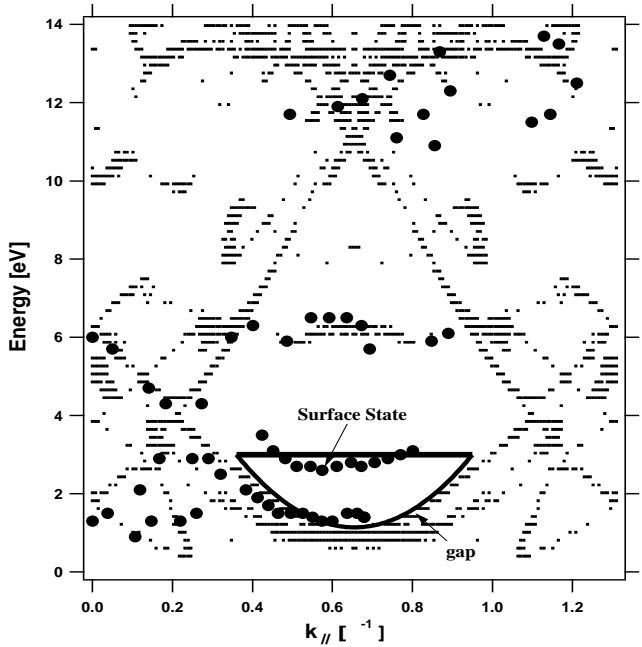,width=9cm,height=9cm}}
\caption{\label{Fig. 10.} Experimental resonances ($\bullet$) and
bulk predicted transitions (black squares). The surface state is
clearly contained in the gap.
 }
\end{figure}
In fact this is in contrast with the dispersion of $S$ which
clearly follows the symmetry of the nearby energy bands,
distinguishing it from an image charge type states which has its
minimum energy at $(\bar{\Gamma})$.  The surface state $S$ has its
minimum energy at the zone boundary ($k_{//} = 0.66$ \AA$^{-1}$).
\section{Conclusions}
We have studied using IPS the empty electronic states of Au(113)
from the Fermi level up to 15 $eV$ along the two main
crystallographic axis: $[\bar{1}10]$ and $[33\bar{2}]$. In
addition we calculated from first principles the Au band
structure.  We used this information to perform a surface
projection of the bulk electronic structure and determine the
locations of the surface energy gaps.  Comparing the experimental
results with the transitions predicted by our calculation we were
able to recognize several surface resonances.  Some of them were
derived from bulk states.  From the calculations we gained insight
on the nature and origin of two surface resonances (along
$[\bar{1}10]$ direction with energies 4.3 $eV$ and 2.7 $eV$ near
normal incidence) and a surface state (along $[33\bar{2}]$, with a
minimum energy of 2.7 $eV$ at $ k_{//} \approx 0.6$\AA$^{-1}$).
Undoubtedly our experimental results in conjunction with the
theoretical analysis presented here shows that isochromat IPS is a
fairly powerful technique to study the unoccupied energy bands of
single crystalline structures.
\section*{Acknowledgment}
We give special thanks to Dr. Dave Zehner for his help in preparing
the crystal.  This research received financial support from
FONDECYT grants \# 1990812 and 1990304 ,Fundaci\'{o}n Andes grant
C-10810/2 and ICM P99-135-F.

\end{document}